\documentclass[]{aastex631}

\usepackage{amsmath}

\begin{document}

\title{Experimental Verification of a One-Dimensional Diffraction-Limit Coronagraph}

\author[0000-0003-2690-7092]{Satoshi Itoh}
\author{Taro Matsuo}
\author{Shunsuke Ota}
\author{Kensuke Hara}
\affiliation{Department of Particle and Astrophysics\\ Graduate School of Science\\ Nagoya University, Furocho, Chikusa-ku, Nagoya, Aichi, 466-8602, Japan}
\author{Yuji Ikeda}
\author{Reiki Kojima}
\affiliation{Laboratory of Infrared High-resolution Spectroscopy\\ Koyama Astronomical Observatory\\ Kyoto Sangyo University, Motoyama, Kamigamo, Kita-ku, Kyoto 603-8555, Japan}
\author{Toru Yamada}
\affiliation{Institute of Space and Astronautical Science\\ Japan Aerospace Exploration Agency\\ 3-1-1, Yoshinodai, Chuou-ku, Sagamihara, Kanagawa 252-5210, Japan}
\author{Takahiro Sumi}
\affiliation{Faculty of Earth and Space Science\\ Osaka University 1-1, Machikaneyama-cho, Toyonaka, Osaka 560-0043, Japan}



\begin{abstract}
We performed an experimental verification of a coronagraph.
As a result, we confirmed that, at the focal region where the planetary point spread function exists, the coronagraph system mitigates the raw contrast of a star-planet system by at least $1\times10^{-5}$ even for the 1-$\lambda/D$ star-planet separation.
In addition, the verified coronagraph keeps the shapes of the off-axis point spread functions when the setup has the source angular separation of 1$\lambda/D$. 
The low-order wavefront error and the non-zero extinction ratio of the linear polarizer may affect the currently confirmed contrast.
The sharpness of the off-axis point spread function generated by the sub-$\lambda/D$ separated sources is promising for the fiber-based observation of exoplanets. The coupling efficiency with a single mode fiber exceeds 50\% when the angular separation is greater than 3--4$\times 10^{-1}\lambda/D$.
For sub-$\lambda/D$ separated sources, the peak positions (obtained with Gaussian fitting) of the output point spread functions are different from the angular positions of sources; the peak position moved from about $0.8\lambda/D$ to $1.0\lambda/D$ as the angular separation of the light source varies from $0.1\lambda/D$ to $1.0\lambda/D$.
The off-axis throughput including the fiber-coupling efficiency (with respect to no focal plane mask) is about 40\% for 1-$\lambda/D$ separated sources and 10\% for 0.5-$\lambda/D$ separated ones (excluding the factor of the ratio of pupil aperture width and Lyot stop width), where we assumed a linear-polarized-light injection.
In addition,  because this coronagraph can remove point sources on a line in the sky, it has another promising application for high-contrast imaging of exoplanets in binary systems.
\end{abstract}



\section{Introduction}
\begin{figure}[htbp]
    \centering
    \includegraphics[width=0.6\linewidth]{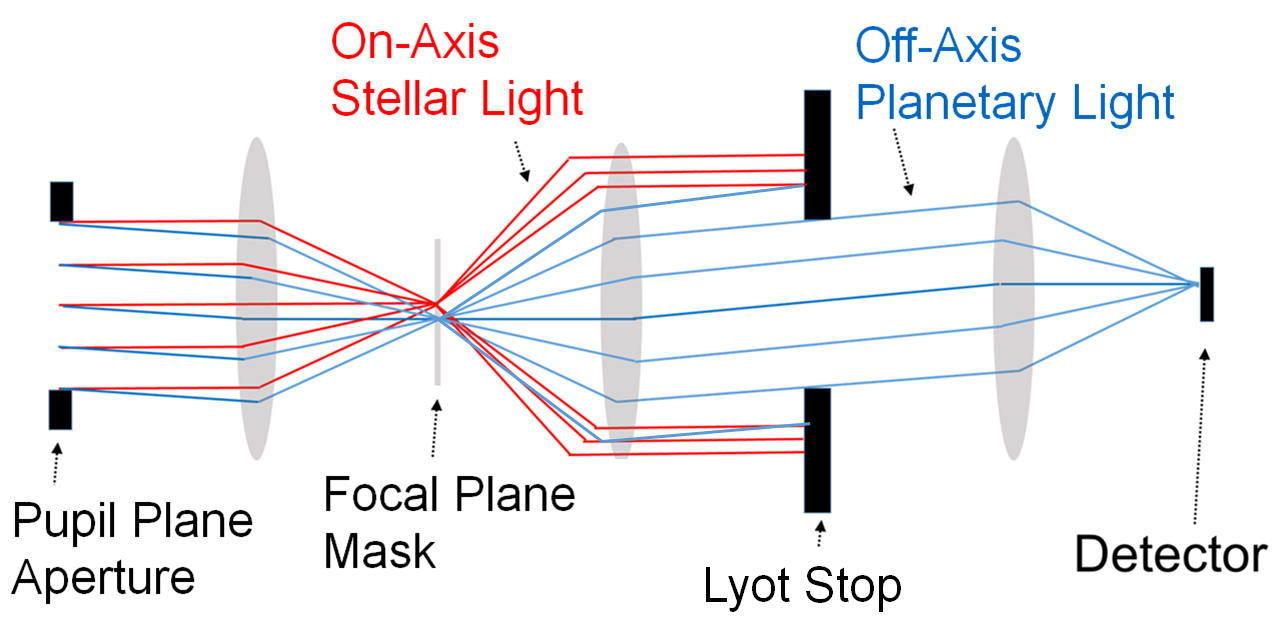}
    \caption{Stellar coronagraphs to detect off-axis planetary light.}
    \label{fig:wic}
\end{figure}

Methods for direct imaging of light from exoplanets contain stellar coronagraphs  \citep{1977Icar...30..422K}, many of which consist of two main elements: focal-plane mask and pupil-plane aperture referred to as Lyot stop (Figure \ref{fig:wic}).
Focal-plane masks selectively modulate (i.e., absorb, reflect, or scatter) the on-axis focused light amplitudes.
This modulation should maintain the transmittance of the off-axis focused light amplitudes to secure a sufficient number of observed photons.
Lyot stops block the stellar light scattered (diffracted) by focal-plane masks to mitigate the contrast between stellar and planetary light.
The other examples observed with steller coronagraphs are circumstellar disks and brown dwarves.   
In addition, future applications of the coronagraph method include the direct imaging and spectrometry of reflected light from Earth-like exoplanets around Solar or Sun-like stars  \citep{2019SPIE11117E..03P,2022JATIS...8c4001J}.

We did not perfectly realize its ``diffraction-limit'' nature (we define the term ``diffraction-limit'' in Appendix A) then, but our past publication  \citep{Itoh+2020} reported a specific theory to construct diffraction-limit coronagraphs strictly.
Although coronagraph masks proposed so far have a wide variety \citep[e.g.,][]{1997PASP..109..815R,2001PASP..113..436B,2002ApJ...570..900K,2004EAS....12..185B,2005ApJ...630..631O,2005ApJ...633.1191M,2005OptL...30.3308F,2006dies.conf..485R,2008PASP..120.1112M,2009OExpr..1720515C,2017OExpr..25.7273B},  only a few types of coronagraph can even approximately achieve their ``diffraction-limit'' performance.
Approximately diffraction-limit coronagraphs have the examples of phase-induced amplitude apodization complex mask coronagraph  \citep{2012SPIE.8442E..4VG} and achromatic interfero-coronagraph  \citep{2006dies.conf..485R}.
The paper   \citep{Itoh+2020} mentioned above states only the one-dimensional diffraction-limit coronagraphs that can null point sources in a line on the sky for (segmented or partially shielded) rectangular pupils; however, the similar theoretical procedure works for diffraction-limit coronagraphs for arbitrary pupils. 
Nevertheless, we have focused on the one-dimensional diffraction-limit coronagraphs because we appreciate their potential to develop into a coronagraph with the fourth-order sensitivity to tilt aberration (i.e., the stellar angular diameters and telescope pointing accuracy).
We can obtain the coronagraph with the fourth-order null  \citep{Itoh+2020} by multiplying the mask function of the one-dimensional ($x$-directional) diffraction-limit coronagraph by that of the one-dimensional ($y$-directional) diffraction-limit coronagraph.
We have also investigated the wide-spectral-band diffraction-limit coronagraphs 
  \citep{Matsuo+2021,2022AJ....163..279I}.
We compiled mathematical explanations of versions of the diffraction-limit coronagraphs other than the one-dimensional ones in Appendix B.

This study shows the result of the first experimental verification of the one-dimensional diffraction-limit coronagraphs.
Section 2 briefly reviews the theory for the one-dimensional diffraction-limit coronagraphs.
In Section 3, we explain the setup of the experiment.
Section 4 evaluates the resultant coronagraphic performance achieved in this experiment.
Section 5 discusses the prospects for applications of the diffraction-limit coronagraphs and possible coronagraphs of the same types as the one-dimensional diffraction-limit coronagraphs.
In Section 6, we summarize the conclusion of the present study.
\section{Theory}

In this section, we first describe the general theory of coronagraphs of rectangular pupils.
Then, using the theoretical framework, we describe the theoretical properties of one-dimensional diffraction-limited coronagraphs.

\subsection{General Theory of Coronagraphs for Rectangular Pupils}
Hereafter, $\vec{\alpha}=\left(\alpha,\beta\right)$ and $\vec{x}=\left(x,y\right)$ respectively indicate the pupil- and focal- coordinates.
The normalization factors $D$ and $f\lambda/D$ normalize these coordinates, where $f$ denotes the effective focal length of the imaging optics, $D$ expresses the full width of the rectangular pupil aperture of the telescope, and $\lambda$ means the observation wavelength.
Since only $D$ and $f\lambda/D$ scale with the magnification ratio of the imaging optics, the normalized coordinates $\vec{\alpha}$ and $\vec{x}$ are independent of the magnification ratio.

In this review, we assume the following rectangular pupil function $P(\vec{\alpha})$:
\begin{equation}
    P\left(\vec{\alpha}\right)=\mathrm{rect}\left[\alpha\right]\mathrm{rect}\left[\beta\right].\footnote{\begin{equation}
\mathrm{rect}\left[X\right]=\left\lbrace\begin{matrix}
1&\left(\left|X\right|<1/2\right)\\
1/2&\left(\left|X\right|=1/2\right)\\
0&\left(\left|X\right|>1/2\right)
\end{matrix}\right..
\end{equation}}
\end{equation}
Here, we consider the following functions that can have non-zero values only within the pupil aperture:
\begin{equation}
 b_{kl}\left(\vec{\alpha}\right)=P\left(\vec{\alpha}\right)e^{2\pi i \left(k \alpha +l\beta\right)}\ \ \ \left(k,l\in\mathbb{Z}\right).\label{d}
\end{equation}
The above functions have the following property (orthonormality):
\begin{equation}
    \int _{-\infty}^{\infty}\int _{-\infty}^{\infty} d\alpha  d\beta\  b_{kl}\left(\vec{\alpha}\right)b_{mn}\left(\vec{\alpha}\right)=\delta_{km}\delta_{ln},
\end{equation}
where $\delta_{ij}$ denotes the Kronecker delta.
When $f\left(\vec{\alpha}\right)$ denotes an arbitrary function that can have non-zero values only within the pupil aperture, the following relation holds:
\begin{equation}
    f\left(\vec{\alpha}\right)=\sum_{k=-\infty}^{\infty}\sum_{l=-\infty}^{\infty}w_{kl}b_{kl}\left(\vec{\alpha}\right),
\end{equation}
where
\begin{equation}
    w_{kl}= \int _{-\infty}^{\infty}\int _{-\infty}^{\infty} d\alpha  d\beta\ b_{kl}\left(\vec{\alpha}\right) f\left(\vec{\alpha}\right).
\end{equation}
Hence, we can express $f\left(\vec{\alpha}\right)$ as a linear conbination of the functions $b_{kl}\left(\vec{\alpha}\right)$ with the weights $w_{kl}$.
In other terms, the functions $b_{kl}\left(\vec{\alpha}\right)$ work as complete orthonormal base functions within the pupil aperture.

Fourier transform preserves the inner products of two functions.
Thus, Fourier transform yields focal-plane orthonormal base functions associated with $b_{kl}\left(\vec{\alpha}\right)$ through one-to-one correspondences. 
Fourier transforming Equation (\ref{d}) leads to the following functions on the image plane:
\begin{equation}
\widetilde{b_{kl}}\left(\vec{x}\right)=\mathrm{sinc}\left(x-k\right)\mathrm{sinc}\left(y-l\right)\ \ \ \left(k,l\in\mathbb{Z}\right),\label{eq7}
\end{equation}
where 
\begin{equation}
    \mathrm{sinc}\left(z\right)=\frac{\sin{\pi z}}{\pi z}.
\end{equation}
When $g\left(\vec{x}\right)$ indicates an arbitrary function with a spectral band limited by the pupil aperture $P\left(\vec{\alpha}\right)$, the following relation holds:
\begin{equation}
    g\left(\vec{x}\right)=\sum_{k=-\infty}^{\infty}\sum_{l=-\infty}^{\infty}w'_{kl}\widetilde{b_{kl}}\left(\vec{x}\right),
\end{equation}
where
\begin{equation}
    w'_{kl}= \int _{-\infty}^{\infty} dx \int _{-\infty}^{\infty} dy\ \widetilde{b_{kl}}\left(\vec{x}\right) g\left(\vec{x}\right).
\end{equation}
Hence, the functions $\widetilde{b_{kl}}\left(\vec{x}\right) $ work as complete orthonormal base functions for an arbitrary focal amplitude with its Fourier conjugate limited by the pupil aperture $P\left(\vec{\alpha}\right)$.
The set of base functions $\widetilde{b_{kl}}\left(\vec{x}\right)$ underlie the Nyquist-Shannon sampling theorem  \citep{1949IEEEP..37...10S}.

When $g\left(\vec{x}\right)$ expresses the fourier conjugate of $f\left(\vec{\alpha}\right)$, the weights $w'_{kl}$ and $w_{kl}$ match.
In this case, the weight $w_{kl}$ contains complete information on the complex amplitudes on the pupil and image plane before the coronagraph system. 
More importantly, we can completely express the nature of coronagraph systems as the following linear filtering systems that affect the above weights $w_{kl}$ (we write weights before and after the coronagraph as $w_{kl}^{\mathrm{before}}$ and $w_{kl}^{\mathrm{after}}$, respectively):
\begin{equation}
    w_{mn}^{\mathrm{after}}= \sum_{k=-\infty}^{\infty}\sum_{l=-\infty}^{\infty} A_{klmn} w_{kl}^{\mathrm{before}},
\end{equation}
where the filter tensor $A_{klmn}$ satisfies the following condition:
\begin{equation}
 |A_{klmn}|\leq 1.
\end{equation}
\subsection{One-Dimentional Diffraction-Limit Coronagraph}
\begin{figure}[htbp]
    \centering
    \includegraphics[width=0.9\linewidth]{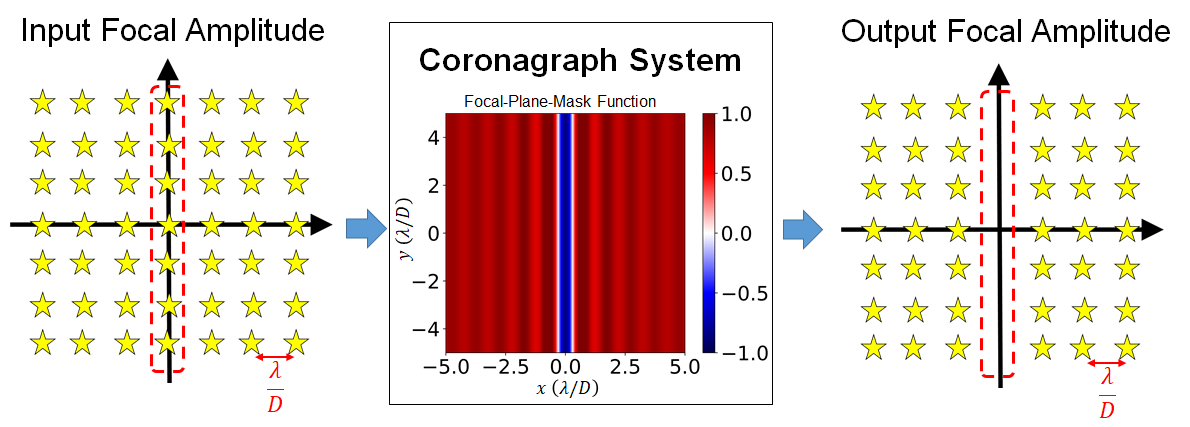}
    \caption{Schematics for how the diffraction-limit coronagraph system works. The left panel expresses focal amplitudes before the coronagraph system. The right panel shows focal amplitudes after the coronagraph system. The yellow stars conceptually indicate the location of the peaks of the orthonormal complete base functions $\widetilde{b_{kl}}\left(\vec{x}\right)$ of Eqation (\ref{eq7}).  The center panel means the coronagraph system and contains the colormap that indicates the focal plane mask function.  The yellow stars encircled with the red dashed curve in the left panel vanished on the right. The erased stars show the base functions with their components perfectly nulled by the coronagraph system. The coronagraph system multiplies the components of the base functions remaining in the right panel by the global transmittance factor $a$. }
    \label{CONS}
\end{figure}

To introduce the coronagraph   \citep{Itoh+2020} to be verified in this study, we assume the following mask function $M\left(\vec{x}\right)$ and Lyot-stop aperture function $L\left(\vec{\alpha}\right)$:
\begin{equation}
    M\left(\vec{x}\right)=a\left(1-2\mathrm{sinc}\left(2x\right)\right),\label{eq5}
\end{equation}
\begin{equation}
    L\left(\vec{\alpha}\right)=P\left(\vec{\alpha}\right),
\end{equation}
where $a=0.697...$; the constant $a$ normirizes $M\left(\vec{x}\right)$ so that $\left|M\left(\vec{x}\right)\right|\leq 1$.
Although we adopted the conservative assumption that $\left|M\left(\vec{x}\right)\right|\leq 1$, this comes not from physical principles. For example, using the phase-induced amplitude apodization \citep{Guyon+2010} enables mask-function values above one or below minus one.
We show a colormap of the values of the mask function $M\left(\vec{x}\right)$ in the center panel of Figure \ref{CONS}.
When we assume a base function $b_{kl}\left(\vec{\alpha}\right)$ as the pupil amplitude, we obtain the following output amplitude on the Lyot-stop ($\mathcal{F}[...]$ means Fourier conjugates):
\begin{equation}
L\left(\vec{\alpha}\right)\mathcal{F}\left[M\left(\vec{x}\right)\mathcal{F}\left[b_{kl}\left(\vec{\alpha}\right)\right]\right]=a\left(1-\delta_{0k}\right)b_{kl}\left(\vec{\alpha}\right).
\end{equation}
Hence, we can express the filter tensor of this coronagraph as follows:
\begin{equation}
    A_{klmn}=a\left(1-\delta_{0k}\right)\delta_{km}\delta_{ln}.\label{eq16}
\end{equation}
The factor $\delta_{km}\delta_{ln}$ in Equation (\ref{eq16}) means identity transform.
The constant factor $a$ of the mask function degrades the global throughput of the system.
Only the factor $1-\delta_{0k}$ expresses the relative throughputs of the weights $w_{kl}$.
The relative-throughput factor $1-\delta_{0k}$ takes $0$ at $k=0$ and $1$ at $k\neq0$, showing the diffraction-limit (theoritical-limit) performance along the $x$-direction.
In addition, this coronagraph can remove all sources distributed on a one-dimensional line: $x=0$.
Hence, we refer to this coronagraph as the one-dimensional diffraction-limit coronagraph.
We schematically show how the one-dimensional diffraction-limit coronagraph works in Figure \ref{CONS}.

\section{Experiment}
We performed the following experiment to verify the concept of the above one-dimensional diffraction-limit coronagraph system. 
\subsection{System Overview}
\begin{figure}[htbp]
    \centering
    \includegraphics[width=\linewidth]{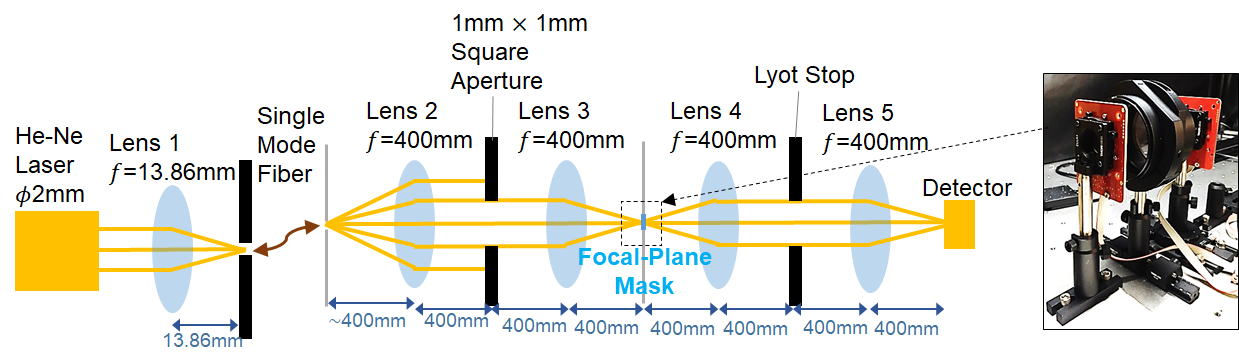}
    \caption{Schematics for the experimental setup.
The right panel is the picture of the focal-plane-mask unit used in the experiment. The focal-plane-mask unit consists of a half-wave plate with a spatially varying fast-axis-direction pattern between two linear polarizers. The light paths after the square aperture in this figure are only conceptual because they ignore the diffraction due to the small Fresnel number of the imaging system. Nevertheless, even when we consider the diffraction effects with the approximation level of the Fresnel diffraction (paraxial approximation), the positions of the pupil and focal planes match those in this figure, and the amplitude on the focal plane corresponds to the Fourier conjugate of the pupil amplitude. Hence, this configuration is valid for experimental purposes. However, real instruments may need to have larger pupil diameters (e.g., 10mm) and focal-plane masks with a finer scale compared to this configuration.}
    \label{fig:my_label_0}
\end{figure}
Figure 1 indicates schematics for the experimental setup. 
The output of the single-mode fiber (Thorlabs, P1-630A-FC-2) with a mode field diameters of 3.6--5.3$\mathrm{\mu m}$ (catalogue value) simulates the stellar quasi-point source.
The square aperture ($D=1\mathrm{mm}$) mimics the telescope aperture.
Our numerical simulation shows that this spread width of the coherent light source\footnote{A coherent light source means a spread light source such that arbitrary pairs of point sources in the light source have spatial coherence (correlation).  The coherent light source does not simulate stars perfectly because stars are perfectly incoherent light sources practically.} brings an amplitude non-uniformity of about $1\times 10^{-3}$ (relative, root mean square) on the rectangular pupil aperture. 
We set the F-number $f/D$ of the focusing system to 400 so that the surface shape of the lenses less affects the wavefront aberrations.
We used a light source He-Ne laser with a wavelength $\lambda$ of 632.8nm.
A preliminary experiment with a grating spectrometer showed that the spectral broadness and the central-wavelength shift from 632.8nm are less than 0.45nm ($7.0\times10^{-4} \lambda$).
The focal-plane normarization factor $f\lambda/D$ is $2.53\times 10^{-1}\mathrm{mm}$.
As the detector, we used a CCD camera (SBIG, ST-402ME) with $765\times510$ pixels (9-$\mathrm{\mu m}$ pixel pitch).
All the lenses except Lens 3, just before the focal-plane mask, are ready-made products (Edmund Optics, \#88595) with the focal-length tolerance of $\pm 2\%$. 
On the other hand, Lens 3 satisfies the strict requirement of the focal-length tolerance $\pm 0.1\%$ (custom-made by Natsume Optical Corporation).
We set the Lyot-stop width to about 0.75mm. 
If the clear aperture of the focal-plane mask spans an infinite area, the Lyot-stop size can be 1mm (the same size as the first pupil aperture) in principle. In this experiment, the clear aperture of the focal-plane mask has a diameter of about 25.4mm (about 100$\lambda/D$). This window function on the focal plane (the effect of the finite area of the focal-plane mask) acts as a low-pass filter to the pupil amplitude before the light suppression by the Lyot stop, resulting in some stellar leak within the Lyot-stop aperture.  Adding a focal-plane apodizing that soothes the edge of the focal-plane window function in future work leads to the suppression of the leak within the Lyot stop. In this case, we expect that we can broaden the Lyot-stop aperture compared to this experiment.
\begin{figure}[htbp]
    \centering
    \includegraphics[width=0.7\linewidth]{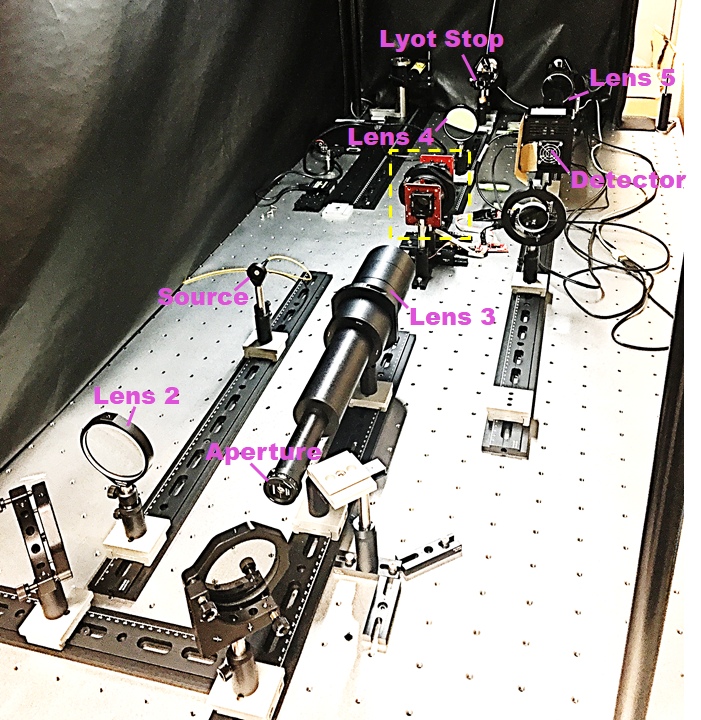}
    \caption{Picture for the experimental setup after the single-mode fiber. The several plane mirrors fold the optical axis to resolve the size limit of the used optical bench. The part involved in the yellow dashed rectangular is the focal-plane-mask unit. }
    \label{fig:pic}
\end{figure}
We show the picture of the whole setup in Figure \ref{fig:pic}.
\subsection{Focal-Plane-Mask Unit}
To implement the mask function $M\left(\vec{x}\right)$ of Equation (\ref{eq5}), we used a half-wave plate with a spatially varying fast-axis-direction pattern (custom-made by Beam Co.) located between two linear polarizers (Figure \ref{fig:my_label_0}).
This configuration can simultaneously modulate the amplitude and $\pi$-radian phase (plus/minus sign) of the focused light-wave complex amplitude so that the mask function takes the real-number values from -1 to 1.
In the present experiment, we used the following design of the spatial pattern of the fast-axis angles $\theta\left(\vec{x}\right)$:
\begin{equation}
 \theta\left(\vec{x}\right)=\frac{\arcsin{\left(M\left(\vec{x}\right)\right)}}{2}.
\end{equation}

We can observe how the above configuration works theoretically by evaluating the total Jones matrix that acts to the Jones vector $\left(
E_x\left(\vec{x}\right)\ E_y\left(\vec{x}\right)\right)^{T}$ that consists of the components of the  x- and y- directional electric fields at the position $\vec{x}$ as follows:
\begin{equation}
\begin{pmatrix}
0&0\\
0&1
\end{pmatrix}
\hat{R}\left(-\theta\left(\vec{x}\right)\right)
\begin{pmatrix}
-1+\delta& 0\\
0 & 1
\end{pmatrix}
\hat{R}\left(\theta\left(\vec{x}\right)\right)
\begin{pmatrix}
1&0\\
0&0
\end{pmatrix}
=\left(1-\frac{\delta}{2}\right)\begin{pmatrix}
0&0 \\
\sin{\left(2\theta\left(\vec{x}\right)\right)}&0
\end{pmatrix},\label{eq8}
\end{equation}
where
\begin{equation}
\hat{R}\left(\gamma\right)=
\begin{pmatrix}
\cos{\gamma} & \sin{\gamma}\\
-\sin{\gamma} & \cos{\gamma}
\end{pmatrix},
\end{equation}
and $\theta\left(\vec{x}\right)$ denotes the fast-axis angles of the half-wave plate relative to the x-direction; the symbol $\delta$ expresses the error of modulation at the half-wave plate. 
The error $\delta$ (such as the phase error caused by the difference between the design and observation wavelengths) can take complex-number values and $\delta=0$ in the case without error.
To see how the focal-plane-mask unit works, we multiply the Jones matrix of Equation (\ref{eq8}) to the Jones vector $\left(
E_x\left(\vec{x}\right)\ 0\right)^{T}$ of a perfectly x-polarized incident beam as the following:
\begin{equation}
\left(1-\frac{\delta}{2}\right)\begin{pmatrix}
0&0 \\
\sin{\left(2\theta\left(\vec{x}\right)\right)}&0
\end{pmatrix}
\begin{pmatrix}
E_x\left(\vec{x}\right) \\
0
\end{pmatrix}=\begin{pmatrix}
0 \\
\left(1-\frac{\delta}{2}\right)\sin{\left(2\theta\left(\vec{x}\right)\right)}E_x\left(\vec{x}\right)
\end{pmatrix}\label{eq11}
\end{equation}
When the fast-axis angle $\theta\left(\vec{x}\right)$  varies from $-\frac{\pi }{4}$ to $\frac{\pi }{4}$, the factor $\sin{\left(2\theta\left(\vec{x}\right)\right)}$ in Equation (\ref{eq11}) continuously varies from $-1$ to $1$.
Hence, for a perfectly x-polarized incident beam, spatially variant fast-axis angles $\theta\left(\vec{x}\right)$ leads to the mask function taking real numbers from $-\left(1-\frac{\delta}{2}\right)$ to $\left(1-\frac{\delta}{2}\right)$.
Note that the output beam shows a perfect y-polarization. 
The error factor $\left(1-\frac{\delta}{2}\right)$ changes only the global transmittance of the mask (the mask transmittance independent of the focal plane coordinates $\vec{x}$).
Thus, the error factor $\left(1-\frac{\delta}{2}\right)$ does not affect the coronagraphic performance other than the global throughput.
Hence, the focal-plane-mask unit has an achromatic nature in principle.
This implementation method of the focal-plane mask assumes a perfectly linear-polarized-light injection. Hence, from the perspective of the high throughput (in the case of unpolarized-light injection), other methods for the implementation of the mask may work more effectively.
\subsection{Data Acquisition}
\begin{figure}[htbp]
    \centering
    \includegraphics[width=0.8\linewidth]{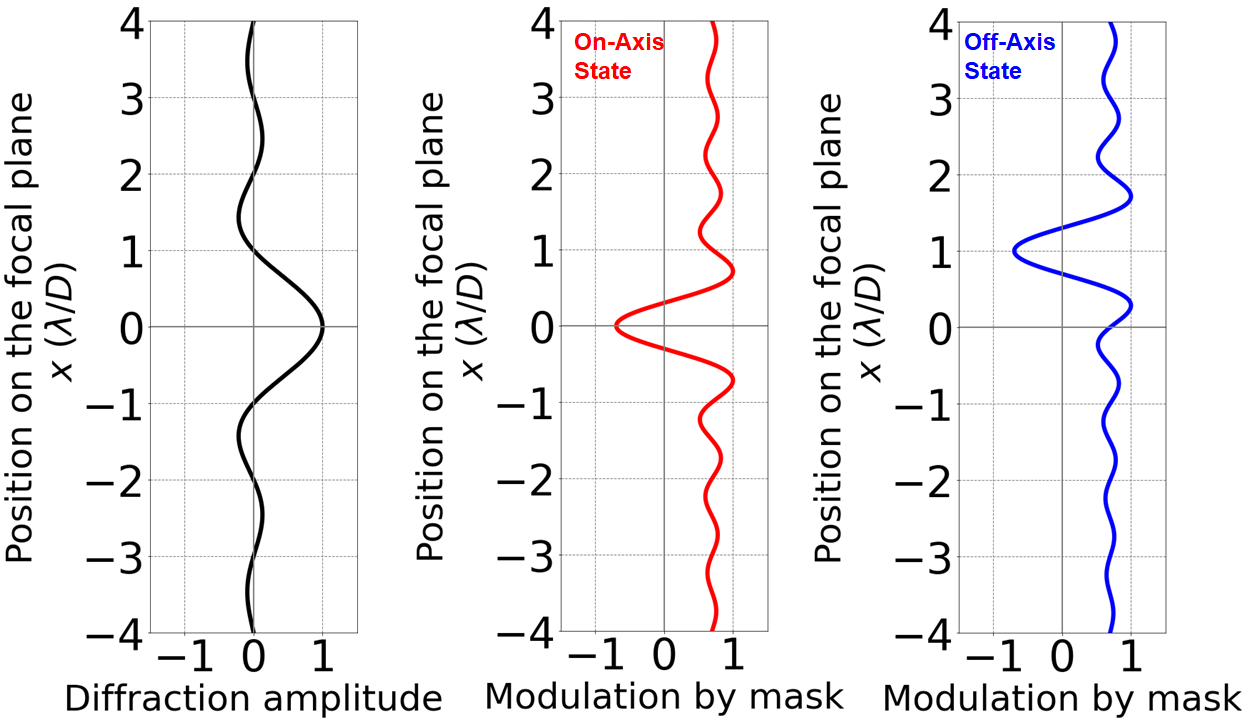}
    \caption{The definition of on- and off-axis states. The difference between the two states derives from the difference in the $x$-position of the center of the focal plane mask. The black curve indicates the focal diffraction amplitude function to be modulated by the focal plane mask. The red and blue curves show the mask function of the on- and off-axis state, respectively.} 
    \label{fig:my_label_onfax}
\end{figure}

To evaluate the achieved contrast-reduction performance, we define the following two states of the experimental setup.
 We refer to them as the on-axis state and the off-axis state.
The difference between the two states comes from the lateral position of the half-wave plate in the focal-plane-mask unit (Figure \ref{fig:my_label_onfax}).
In the on-axis state, the center of the half-wave plate matches the center of the diffraction amplitude profile.
The on-axis state simulates the state with the source angular separation of $0\lambda/D$.
We defined the off-axis state as the state with the lateral position of the half-wave plate where the intensity peak on the detector shows the first local maximum value after the setup leaves the on-axis state.
The off-axis state approximates the state with the source angular separation of $1\lambda/D$.

To secure the high dynamic range of the measurement, we obtained the data with the different exposure times for the two states: 100s for the on-axis state and 0.04s for the off-axis state.
We also obtained dark images for the two types of integration times and subtracted them from the exposed images.
We execute the 3$\times$3 binning to all the acquired image data.
Thus, the resultant data arrays have a size of 255$\times$170. 
\section{Result and Consideration}
\subsection{Result}
\begin{figure}[htbp]
    \centering
    \includegraphics[width=1.0\linewidth]{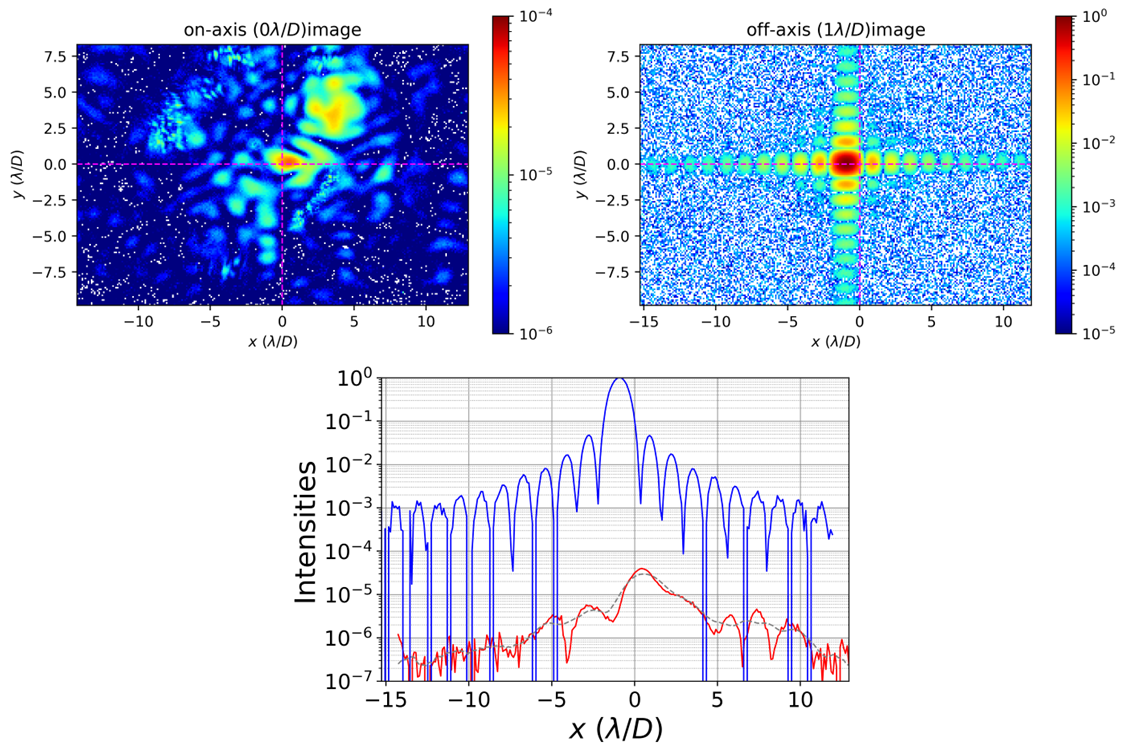}
    \caption{The verification result of the contrast-reduction performance: (top left) the image data (255 $\times$ 170) acquired with the on-axis state (angular separation of 0$\lambda/D$), (top right) the data for the off-axis state (angular separation of 1$\lambda/D$), and (bottom) the comparison of the two data (the red and blue curves for the on- and off-axis state, respectively) along the dashed magenta horizontal lines of the top panels. The gray dashed curve in the bottom panel indicates the moving average (kernel width $2\lambda/D$) of the red curve. The dashed magenta vertical lines in the top panels show the position of the focal-plane mask center. We took the focal-plane mask centers as the $x$-axis origins.} The data in all panels indicate the value of ADU/s normalized by the peak value of the image data of the off-axis state. 
    \label{fig:my_label_res}
\end{figure}
We show the verification results of the contrast-reduction performances in Figure \ref{fig:my_label_res}.
From the result, we observe the light leak at around the 1-$\lambda/D$ separation in the on-axis state falls below $1\times10^{-5}$ times the peak value of the off-axis state.
Hence, this experiment demonstrated the ability of this coronagraph to mitigate the contrast at least by $1\times10^{-5}$ even at the 1-$\lambda/D$ separation.
The image of the off-axis state in Figure \ref{fig:my_label_res} shows the nearly completely same point-spread-function profile as the one before affected by the coronagraph system.
We can explain the resultant unmodulated point spread function with the following mathematics for evaluating the output pupil amplitude on the Lyot-stop plane (see also Appendix A):
\begin{eqnarray}
    L\left(\vec{\alpha}\right)\mathcal{F}\left[M\left(\vec{x}\right)\mathrm{sinc}\left(x-1\right)\right]&=&L\left(\vec{\alpha}\right)\mathcal{F}\left[a\left(1-2\mathrm{sinc}\left(2x\right)\right)\mathrm{sinc}\left(x-1\right)\right] \nonumber \\
    &=&aL\left(\vec{\alpha}\right)\mathcal{F}\left[\mathrm{sinc}\left(x-1\right)\right]-aL\left(\vec{\alpha}\right)\left(\mathrm{rect}\left[\frac{\alpha}{2}\right]\ast \left(\mathrm{rect}\left[\alpha\right]e^{2\pi i \left(1\right)\alpha} \right) \right) \nonumber \\
    &=&aL\left(\vec{\alpha}\right)\mathcal{F}\left[\mathrm{sinc}\left(x-1\right)\right]-aL\left(\vec{\alpha}\right)\mathrm{sinc}\left(1\right)\nonumber \\
    &=&aL\left(\vec{\alpha}\right)\mathcal{F}\left[\mathrm{sinc}\left(x-1\right)\right],\label{20}
\end{eqnarray}
where we used the convolution theorem (of the Fourier transform) in operation between the first and second rows, and the symbol $*$ means convolution.

\subsection{Consideration}
\begin{figure}[htbp]
    \centering
    \includegraphics[width=0.6\linewidth]{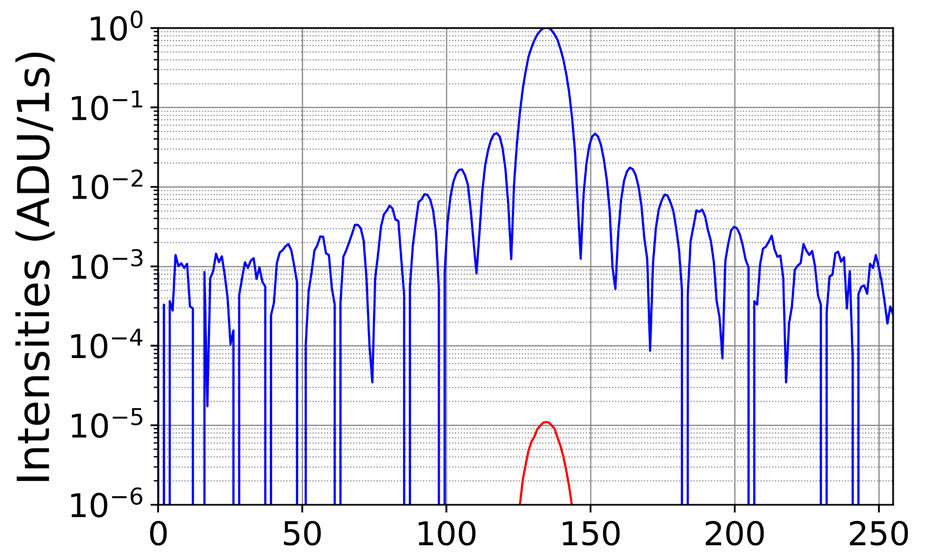}
    \caption{The performance limitation by the non-zero extinction ratio of the linear polarizers. The blue curve indicates the same plot as Figure \ref{fig:my_label_res}. The red curve is data measured after removing the half-wave plate between the two orthogonal linear polarizers from the on-axis state. We have corrected the red curve with the multiplication of the theoretical global throughput factor of the mask, $a^2$.}
    \label{fig:my_label_pollim}
\end{figure}

We consider the following major limiting factors of the current demonstrated performance: (i) the amplitude and phase error of the wavefront that makes the speckles on the focal plane and (ii) the non-zero extinction ratio of the linear polarizers.
We had anticipated the speckles because the experimental setup includes no active wavefront correction instruments.
We had also expected the effect of the non-zero extinction ratio of the linear polarizers because we had measured light leaks from the two orthogonal linear polarizers in advance.
The leak caused by the non-zero extinction ratio corresponds to the contrast of  $1\times10^{-5}$ in its peak value  (Figure \ref{fig:my_label_pollim}).
Hence, to confirm contrast-reduction performance higher than that in this experiment, we need a wavefront-correction instrument and linear polarizers with high extinction performance. 
\section{Discussion}
\subsection{Prospects for Applications}
\subsubsection{Single-Mode-Fiber-Injected High-Contrast Imaging to Detect Exoplanets}
\begin{figure}[htbp]
    \centering
    \includegraphics[width=1.0\linewidth]{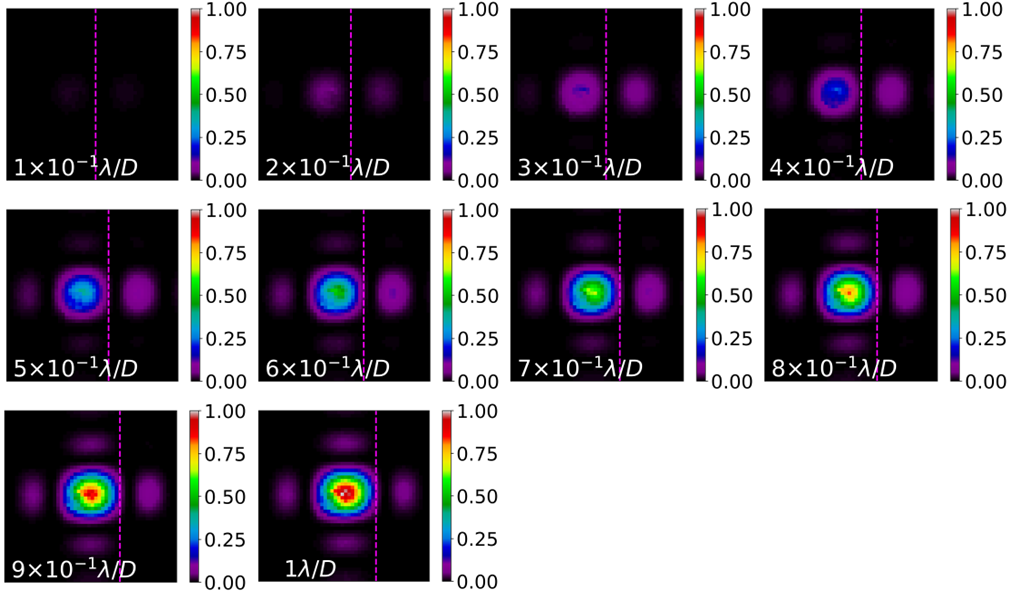}
    \caption{The detected focal intensity in the experiments when the source separation angles are $k\times 10^{-1} \lambda/D$ ($k=1,2,...,10$).  The magenta dashed lines indicate the position of the mask center.  }
    \label{fig:oaps}
\end{figure}

 Figure \ref{fig:oaps} shows observed point spread functions in this experiment at the source separations of 0.1--1$\lambda/D$.
Even for sub-$\lambda/D$ separated source, the one-dimensional diffraction-limit coronagraphs yield the point spread functions with a broadness similar to the one before the coronagraph.  
This feature benefits high-contrast imaging/spectroscopy \citep{2015A&A...576A..59S} using single-mode optical fiber \citep{2017A&A...604A.122J} to detect exoplanets.
A single-mode fiber selectively transmits only one mode of light amplitude and thus may reduce the contrast between stellar and planetary lights.
Similar but different techniques involve the vortex fiber nulling \citep{2019SPIE11117E..16R}.

\begin{figure}[htbp]
    \centering
    \includegraphics[width=1.0\linewidth]{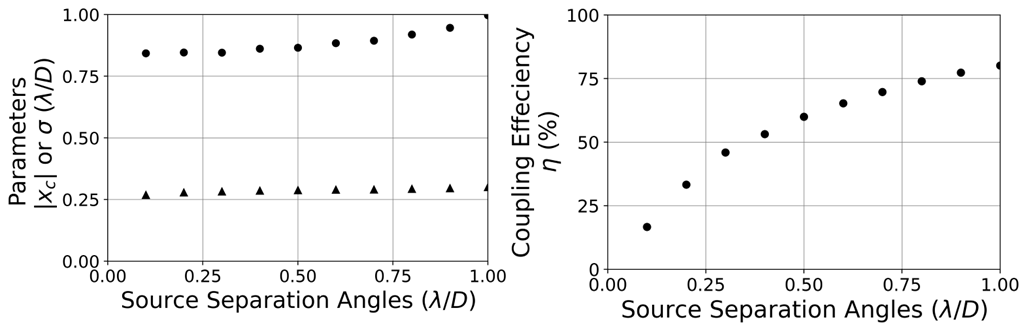}
    \caption{(Left) the absolute values of resultant parameters $|x_c|$ (black filled circles) and $\sigma$ (black filled triangles) of the fitting of the measured point spread functions with the fitting function $\left|g\left(\vec{x}\right)\right|^2$. (Right) the evaluated coupling efficiencies $\eta$ between the fiber-mode function $g\left(\vec{x}\right)$ with the best-fit parameters $(\sigma,x_c,y_c)$ and the amplitude spread function $S\left(\vec{x}\right)$ evaluated by Equation (\ref{kkkk}).} 
    \label{fig:ce}
\end{figure}

We define the coupling efficiency $\eta$ with single-mode fibers of a fiber-mode function $g\left(\vec{x}\right)$  as follows:
\begin{equation}
    \eta= \left|\int_{\mathbb{R}^2}\!\! d\vec{x} \ g\left(\vec{x}\right)^{\ast} S\left(\vec{x}\right)\right|^2, \label{qqqq}
\end{equation}
where we assume normailization of the fiber-mode function $\int_{\mathbb{R}^2}\left|g\left(\vec{x}\right)\right|^2=1$ and the amplitude spread function $\int_{\mathbb{R}^2}\left|S\left(\vec{x}\right)\right|^2=1$.
We evaluated the fiber-coupling efficiency $\eta$ based on the data of the light intensity profile indicated in Figure \ref{fig:oaps} assuming the following fiber-mode function (the square root of the two-dimensional normal distribution):
\begin{equation}
    g\left(\vec{x}\right)=\left(\frac{1}{2\pi \sigma^2}e^{-\frac{\left(x-x_c\right)^2+\left(y-y_c\right)^2}{2\sigma^2}}\right)^{\frac{1}{2}}.\label{poiu}
\end{equation}
First, we determined the parameters $\sigma$ and $\left(x_c, y_c\right)$ by a gaussian fitting of the data in Figure \ref{fig:oaps} with the gaussian fitting function $\left|g\left(\vec{x}\right)\right|^2$; from Equation (\ref{poiu}), $\left|g\left(\vec{x}\right)\right|^2=\frac{1}{2\pi \sigma^2}e^{-\frac{\left(x-x_c\right)^2+\left(y-y_c\right)^2}{2\sigma^2}}$.
Then, we evaluated the coupling efficiencies $\eta$ using the fiber-mode functions with the fitted parameters. 
The left panel of Figure \ref{fig:ce} shows the resultant fitting parameters $x_c$ and $\sigma$.
We can observe from the panel that the $x$-directional center position  $x_c$ and the width $\sigma$ approximately make no variations at the separation angles below $1\lambda/D$.
Although the measured data $\left|S\left(\vec{x}\right)\right|^2$ include no information on the phase of light, from the theoretical perspective, the light amplitudes in the main lobes should have the same phase.
Thus, we assumed the following in the evaluation of the coupling efficiencies $\eta$ in Equation (\ref{qqqq}):
\begin{equation}
    S\left(\vec{x}\right)=\sqrt{\left|S\left(\vec{x}\right)\right|^2}.\label{kkkk}
\end{equation}
Appendix C assesses the validity of the above method of evaluating coupling efficiency.
We show the resultant value of the coupling efficiencies $\eta$ in the right panel of Figure \ref{fig:ce}.
The resultant coupling efficiency $\eta$ exceeds 75\% when the source separation angles are larger than $8\times10^{-1}\lambda/D$ and exceeds 50\% when the source separation angles are larger than 3--4$\times10^{-1}\lambda/D$.

\begin{figure}[htbp]
    \centering
    \includegraphics[width=0.6\linewidth]{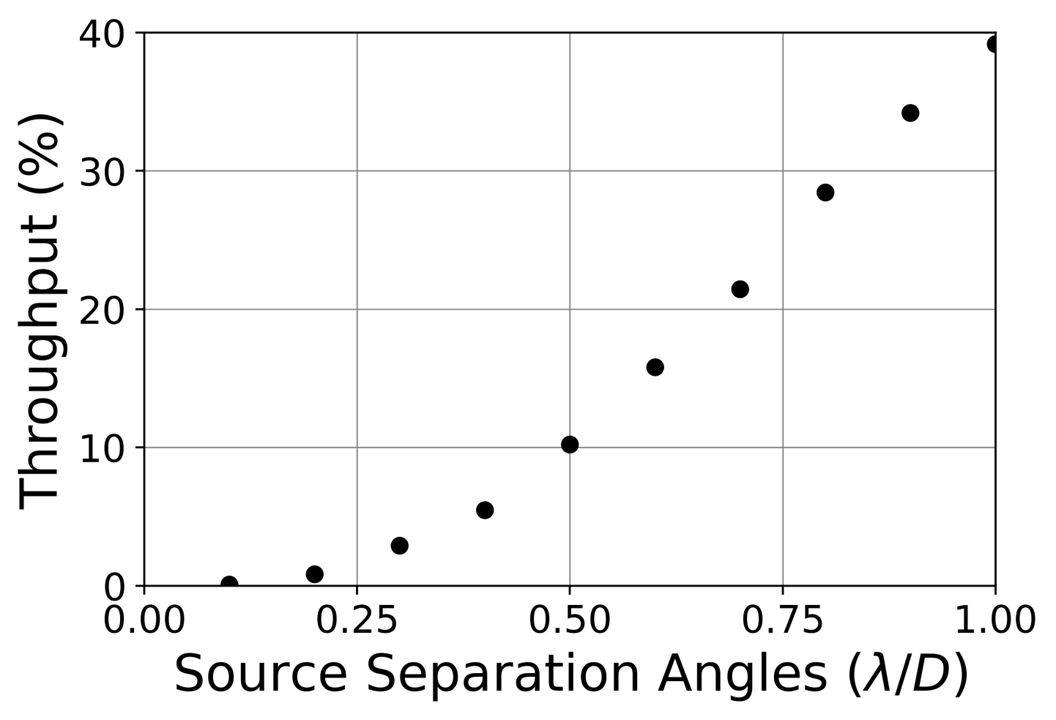}
    \caption{The off-axis throughputs (with respect to no focal plane masks) for different source angular separations. These throughputs include the fiber coupling efficiency in Figure \ref{fig:ce} and exclude the factor of the ratio of pupil aperture width and Lyot stop width; we assumed a perfectly linear-polarized-light injection.}
    \label{fig:tp}
\end{figure}

We can evaluate the off-axis throughput (with respect to no focal plane masks) by multiplying the following three factors: (1) the constant factor ($a^2=0.486$) of the mask function, (2) the peak values of the point spread functions (in Figure \ref{fig:oaps}) normalized by the one in the case with 1 $\lambda/D$, and (3) the coupling efficiency $\eta$. Figure \ref{fig:tp} shows the result of the evaluation.
In this experiment, we implemented the focal plane mask in a particular way that requires perfectly linear-polarized incidence. Hence, the values in Figure \ref{fig:tp} also assumes perfectly linear-polarized incidence. When we use other implementation methods, we do not always need the assumption of a perfect linear polarization incidence. The constant factor ($a^2=0.486$) of the mask function comes from the conservative assumption on the mask function: $\left|M\left(\vec{x}\right)\right|\leq 1$. The phase-induced amplitude apodization may provide the focal-plane-mask function with values greater than one or less than minus one, leading to higher throughput than the values in Figure \ref{fig:tp}.

\subsubsection{High-Contrast Imaging of Exoplanets in Binary Systems}
The one-dimensional diffraction-limit coronagraph can remove the point sources over a line in the sky.
A line can involve arbitrary two points on the two-dimensional plane.
Thus, the one-dimensional diffraction-limit coronagraph can hide binary stars (two quasi-point sources) that host their planets with the diffraction-limit performance. 
Binary stars exist ubiquitously \citep{2010ApJS..190....1R,2015MNRAS.449.2618W}, and many known planets orbit in binary systems  \citep{2022ApJ...935..141S}.
Since the high-contrast direct imaging of the planet can directly detect the light quantum reflected or emitted from the planets, it can yield the spectral characterization of the light quantum from the planets to help us to understand the surface environments of exoplanets. 
If possible companions need to share the orbital plane with the binary systems to have stable orbits, the one-dimensional diffraction-limit coronagraph cannot work to observe the edge-on companions around the binary systems that are good targets for radial-velocity-related observation because these targets fall in the least-throughput region of the one-dimensional diffraction-limit coronagraph.
\section{Conclusion}
In this study, we experimentally verified a new type of coronagraph that should completely preserve the shape of the off-axis point spread function when the source separation angle is $1\lambda/D$ and approximately preserves it for source separation angles below $\lambda/D$.
The focal-plane mask used in the experiment has the mask-function value of the continuous real value from -1 to 1. 
The mask is a half-wave plate with spatially variant fast-axis orientation inserted between two linear polarizers. 
The result showed that the coronagraph system reduced the raw contrast of a star-planet system by $1\times10^{-5}$ even for the 1-$\lambda/D$ star-planet separation.
We also find that, even for the 1-$\lambda/D$ star-planet separation, the final point spread function has the same profile as one before passing through the coronagraph system.
We believe that the wavefront error and a non-zero extinction ratio of the linear polarizers limited the currently confirmed performance. 
Hence, testing more accurate raw contrast performances requires a wavefront-correction instrument and linear polarizers with high extinction performance.
Promising features of this coronagraph involve the sharpness of the sub-$\lambda/D$ off-axis point spread function. 
The peak positions (obtained with Gaussian fitting) of the output point spread functions moved from about $0.8\lambda/D$ to $1.0\lambda/D$ as the angular separation of the light source varies from $0.1\lambda/D$ to $1.0\lambda/D$.
The coupling efficiency with a single mode fiber exceeds 50\% when the angular separation is greater than 3--4$\times 10^{-1}\lambda/D$.
Hence, this coronagraph works for the fiber-based observation of exoplanets. This coronagraph can remove point sources in a line in the sky and thus has another promising application for high-contrast imaging of exoplanets in binary systems.

\section*{Acknowledgement}
The anonymous reviewers have given valuable comments to improve this manuscript. 
Niko Optics Co., Ltd has performed custom-made anti-reflection coatings on the linear polarizers used in the experiment.

\bibliography{main}{}
\bibliographystyle{aasjournal}
\appendix
\section{Diffraction-Limit Coronagraph}
We can understand how a coronagraph works on light amplitudes well using the mathematics of vector spaces with the operation of Hermitian inner products  \citep{2006ApJS..167...81G}.
From the perspectives of mathematics of abstract vector space\footnote{Since two-dimensional Fourier transform preserves the Hermitian inner products between arbitrary two functions, in terms of abstract vector space, function vector space of pupil and focal amplitudes have no difference.
Thus, we can regard them as a single abstract vector space.}, the coronagraph separates the vector space $V$ formed by the entire mathematically realizable light amplitude into a direct sum of two vector subspace as the following: $V=W\oplus W^{\perp}$, where the coronagraph removes the vector space $W$  and transmits the vector space $W^{\perp}$; we refer $V$, $W$, and $W^{\perp}$ as entire space, null subspace, and transmission subspace, respectively. 
The direct sum of vector subspaces $A=B\oplus C$ means that $A=B\cup C$ and that $b\cdot c =0\  \left(\forall b \in B, \ \forall c \in C\right)$, where $b\cdot c$ denotes inner product between two vectors $b$ and $c$.  
The shape of the pupil aperture limits the entire vector space $V$.
For example, the Zerinike Polynomials \citep{1934MNRAS..94..377Z} work as the complete orthogonal base functions of the function vector space formed by the entire vector space $V$ in the case of circular pupils.

Naturally, we can define the term ``diffraction-limit'' coronagraph as a coronagraph with null subspace $W$ formed by linear combinations of amplitudes from some point sources in the sky. 
In other words, for a diffraction-limit coronagraph that removes $N$ point sources, $W=\left\lbrace\left.\sum_{k=1}^{N}s_{k}S\left(\vec{x}-\vec{\theta}_{k}\right)\right|s_k \in \mathbb{C}\right\rbrace$, where $S\left(\vec{x}\right)$ denotes an amplitude spread function (the Fourier conjugate of the pupil function) normalized so that $\int_{\mathbb{R}^2}d\vec{x}|S\left(\vec{x}\right)|^2=1$, the symbol $\vec{x}$ indicates the coordinates on the focal plane, and the symbol $\vec{\theta}_k$ means the angular positions of the $k$-th point sources on the sky.
We can write a focal amplitude $S\left(\vec{x}-\vec{\theta}\right)$ from a point sorce with an arbitrary angular position $\vec{\theta}$ as a sum of the components of null subspace $W$ and transmission subspace $W^{\perp}$:
\begin{equation}
S\left(\vec{x}-\vec{\theta}\right)=S\left(\vec{x}-\vec{\theta}\right)_{\parallel}+S\left(\vec{x}-\vec{\theta}\right)_{\perp}, \label{pp}  
\end{equation}
where $S\left(\vec{x}-\vec{\theta}\right)_{\parallel}\in W$ and $S\left(\vec{x}-\vec{\theta}\right)_{\perp} \in W^{\perp}$.
When the number of the point sources that the coronagraph removes is unity ($N=1$), we can express
the component in the null subspace $W$ using the Hermitian inner product between $S\left(\vec{x}-\vec{\theta}_{1}\right)$ and $S\left(\vec{x}-\vec{\theta}\right)$:
\begin{eqnarray}
    S\left(\vec{x}-\vec{\theta}\right)_{\parallel}&=&\left(\int_{\mathbb{R}^2}d\vec{x}S\left(\vec{x}-\vec{\theta}_{1}\right)^{\ast}S\left(\vec{x}-\vec{\theta}\right) \right)S\left(\vec{x}-\vec{\theta}_{1}\right)\nonumber\\
    &=&\left(\int_{\mathbb{R}^2}d\vec{x}S\left(\vec{x}\right)^{\ast}S\left(\vec{x}-\left(\vec{\theta}-\vec{\theta_{1}}\right)\right)\right)S\left(\vec{x}-\vec{\theta}_{1}\right).\label{qq}
\end{eqnarray}
Since the Equation (\ref{qq}) includes the autocorrelation function of the amplitude spread function $S\left(\vec{x}\right)$, when the pupil function $P\left(\vec{\alpha}\right)$ satisfies the condition\footnote{The condition holds when the values of $\left|P\left(\vec{\alpha}\right)\right|$ take only one or zero.} that $P\left(\vec{\alpha}\right)=\left|P\left(\vec{\alpha}\right)\right|^2$, the following relation holds:
\begin{equation}
 S\left(\vec{x}-\vec{\theta}\right)_{\parallel}=S\left(\vec{\theta}-\vec{\theta_{1}}\right)S\left(\vec{x}-\vec{\theta}_{1}\right).   
\end{equation}
From Equations (\ref{pp}) and (\ref{qq}), we can obtain the following expression for the component in the transmission subspace:
\begin{equation}
    S\left(\vec{x}-\vec{\theta}\right)_{\perp}=S\left(\vec{x}-\vec{\theta}\right)-S\left(\vec{\theta}-\vec{\theta_{1}}\right)S\left(\vec{x}-\vec{\theta}_{1}\right). \label{ww}
\end{equation}
When $\vec{\theta}-\vec{\theta_{1}}$ takes the null point of the amplitude spread function $S\left(\vec{x}\right)$, the second term of Equation (\ref{ww}) takes zero.
Thus, in this case, the original amplitude spread function before the coronagraph transmits the coronagraph system as it is.
\section{Possible Coronagraphs of the Same Type}
The experiment of this study might work as a first step toward the feasibility confirmation of the concepts of possible coronagraphs of the same type as the coronagraph of this verification.
Thus, we compile these concepts below. 
\subsection{Fourth-Order Diffraction-Limit Coronagraph}
Using the mask function of the one-dimensional diffraction-limit coronagraph $M\left(\vec{x}\right)$ in Equation (\ref{eq5}), we can express the mask function $M_{\mathrm{4th}}\left(\vec{x}\right)$ of the fourth-order diffraction-limit coronagraph \citep{Itoh+2020} as the following:
\begin{equation}
    M_{\mathrm{4th}}\left(\vec{x}\right)=M\left(\vec{x}\right)M\left(\vec{x}^{T}\right),
\end{equation}
where $\vec{x}=\left(x,y\right)$ and $\vec{x}^{T}=\left(y,x\right)$.
In the following manner, we can see how changing the mask function from $M\left(\vec{x}\right)$ to  $M_{\mathrm{4th}}\left(\vec{x}\right)$ leads to the fourth-order sensitivity to the tilt aberration.
The first-order approximation of the pupil-plane tilt aberration due to a small angular shift of $\left(\Delta{\theta_x},\Delta{\theta_y}\right)$ has the expression:
\begin{equation}
P\left(\vec{\alpha}\right)e^{2\pi i \Delta{\vec{\theta}}\cdot\vec{\alpha}}= P\left(\vec{\alpha}\right)\left(1+2\pi i \left(\Delta{\theta_x}\alpha+\Delta{\theta_y}\beta\right)\right).\label{qwe}
\end{equation}
The right-hand side of Equation (\ref{qwe}) includes the linear terms of the pupil coordinates $\alpha$ and $\beta$. 
The one-dimensional masks $M\left(\vec{x}\right)$ and $M\left(\vec{x}^{T}\right)$ can completely remove the pupil function dependent on only $\beta$ and  $\alpha$, respectively.
Hence, the mask function $M_{\mathrm{4th}}\left(\vec{x}\right)$ can remove the linear terms in the right-hand side of  Equation (\ref{qwe}) as well as the constant term on the pupil function $P\left(\vec{\alpha}\right)$.
The $k$-th order terms of $\left(\Delta{\theta_x},\Delta{\theta_y}\right)$ in the pupil amplitude cause the $2k$-th order terms in the focal intensity.
Thus, removing the zeroth and first-order terms in the pupil amplitude of Equation (\ref{qwe}) means achieving the fourth-order sensitivity to the tilt aberration.
We obtain the following filter tensor of this coronagraph:
\begin{equation}
    A^{\mathrm{4th}}_{klmn}=a^2\left(1-\delta_{0k}\right)\left(1-\delta_{l0}\right)\delta_{km}\delta_{ln}.\label{eq1646}
\end{equation}
\subsection{Cosine-Modulated One-Dimensional Diffraction-Limit Coronagraph}
The following expression denotes the mask function of the cosine-modulated one-dimensional diffraction-limit coronagraph \citep{2022AJ....163..279I}:
\begin{equation}
    M_{\mathrm{cos}}\left(\vec{x}\right)=a'\cos{\left(\pi x\right)}\frac{M\left(2\vec{x}\right)}{a},
\end{equation}
where $a'=0.899...$; the constant $a'$ normirizes $ M_{\mathrm{cos}}\left(\vec{x}\right)$ so that $| M_{\mathrm{cos}}\left(\vec{x}\right)|\leq 1$.
Fortunately, the constant $a'$ can take a value larger than the constant $a$ in the mask function $M\left(\vec{x}\right)$.
The cosine factor $\cos{\left(\pi x\right)}$ mathematically changes the focal base functions $\widetilde{b_{kl}}\left(\vec{x}\right)$ in Equation (\ref{eq7}) to the following different base functions $\widetilde{b'_{kl}}\left(\vec{x}\right)$:
\begin{equation}
    \widetilde{b'_{kl}}\left(\vec{x}\right)=\left(-1\right)^{k}2\mathrm{sinc}\left(2\left(x-k\right)\right)\mathrm{sinc}\left(y-l\right) \ \ \ \left(k,l\in\mathbb{Z}\right),
\end{equation}
where we used the following mathematical formula:
\begin{equation}
    \cos{\left(\pi x\right)}\mathrm{sinc}\left(x-k\right)=\left(-1\right)^{k}\mathrm{sinc}\left(2\left(x-k\right)\right).
\end{equation}
The factor other than the cosine factor in the mask function works like a one-dimensional diffraction-limit coronagraph for the new base functions $\widetilde{b'_{kl}}\left(\vec{x}\right)$.
The Lyot stop aperture function has the following expression:
\begin{equation}
   L_{\mathrm{cos}} =\mathrm{rect}\left[\frac{\alpha}{2}\right]\mathrm{rect}\left[\beta\right]
\end{equation}
The filter tensor $A^{\mathrm{cos}}_{klmn}$ of this coronagraph is the same as the one-dimensional diffraction-limit coronagraph except for a constant factor:
\begin{equation}
    A^{\mathrm{cos}}_{klmn}=\frac{a'}{a}A_{klmn}.
\end{equation}
The existence of the cosine factor sharpens the output off-axis point spread functions and localizes them near the point where $x$ is an integer.
Hence, this coronagraph may be more promising for fiber-based spectroscopy of exoplanets than the one-dimensional diffraction-limit coronagraph.
\subsection{Spectroscopic Coronagraph}
A diffraction grating on the pupil plane makes a one-dimensional spectrum on the focal plane.
The spectrum consists of the point spread functions of the different wavelengths approximately located in a line.
The one-dimensional diffraction-limit coronagraph can null the sources on a line.
Nevertheless, the point spread functions of the different wavelengths have different scales in the diffraction-limited case.
Therefore, we cannot sufficiently null the spectrum with the one-dimensional diffraction-limit coronagraph.

The concept referred to as the spectroscopic coronagraph \citep{Matsuo+2021} adopts the mask function $M\left(\vec{x}\right)$ of the one-dimensional diffraction-limit coronagraph modified through the projective transformation of the coordinates $\vec{x}$.
The projective transformation can change a square in the focal plane to a trapezoid.
Hence, with the projection transform, we can approximately optimize the scales of the mask pattern to fit the sizes of the point spread functions in the spectrum.
This change in the mask pattern sacrifices the contrast-mitigation ability for each wavelength but enhances the available spectral bandwidth. 
In other words, the spectral bandwidth and the contrast-mitigation ability have a trade-off relationship.

\subsection{Diffraction-Limit Coronagraph for Arbitrary Pupils}
Here, we assume that the pupil has an arbitrary geometry.
Using the area $s$ of the entire pupil, we redefine the pupil diameter $D$ as $D=\sqrt{s}$. 
With the redefined $D$, we redefine the normalizations of the pupil and focal coordinates $\vec{\alpha}$ and $\vec{x}$.
We assume that the pupil function $P\left(\vec{\alpha}\right)$ takes only zero or one except for the boundary of the pupil and that the Lyot stop function $L\left(\vec{\alpha}\right)$ satisfies $L\left(\vec{\alpha}\right)$=$P\left(\vec{\alpha}\right)$. 
We need to define the following function $Q\left(\vec{x}\right)$ to obtain the expression of the mask function of diffraction-limit coronagraph for an arbitrary pupil:
\begin{equation}
    Q\left(\vec{\alpha}\right)=\left\lbrace\begin{matrix}
1&\left( P\left(\vec{\alpha}\right)\ast P\left(\vec{\alpha}\right)\neq 0\right)\\
0&(P\left(\vec{\alpha}\right)\ast P\left(\vec{\alpha}\right)= 0)
\end{matrix}\right.,
\end{equation}
where $P\left(\vec{\alpha}\right)\ast P\left(\vec{\alpha}\right)$ denotes the autocorrelation function of the pupil function $P\left(\vec{\alpha}\right)$.
Using the function $Q\left(\vec{\alpha}\right)$, we can write the mask function of diffraction-limit coronagraph for an arbitrary pupil as the following:
\begin{equation}
    M_{\mathrm{Arb}}\left(\vec{x}\right)=a''\left(1- \widetilde{Q}\left(\vec{x}\right)\right),
\end{equation}
where the constant $a''$ normirizes $  M_{\mathrm{Arb}}\left(\vec{x}\right)$ so that $\left|   M_{\mathrm{Arb}}\left(\vec{x}\right)\right|\leq 1$ and depends on the pupil geometry.

\section{Validity of the Method of Coupling Efficiency Evaluation.}
\subsection{Justification for Assumed Phase of Output Focal Amplitudes}
Using the following mathematical analysis similar to Equation (\ref{20}), we can evaluate the output focal amplitudes of the input sources with $x$-directional separation angles of $\Theta$:
\begin{eqnarray}
    \mathcal{F}\left[L\left(\vec{\alpha}\right)\mathcal{F}\left[M\left(\vec{x}\right)\mathrm{sinc}\left(x-\Theta\right)\right]\right]&=&\mathcal{F}\left[L\left(\vec{\alpha}\right)\mathcal{F}\left[a\left(1-2\mathrm{sinc}\left(2x\right)\right)\mathrm{sinc}\left(x-\Theta\right)\right] \right]\nonumber \\
    &=&\mathcal{F}\left[aL\left(\vec{\alpha}\right)\mathcal{F}\left[\mathrm{sinc}\left(x-\Theta\right)\right]-aL\left(\vec{\alpha}\right)\left(\mathrm{rect}\left[\frac{\alpha}{2}\right]\ast \left(\mathrm{rect}\left[\alpha\right]e^{2\pi i \Theta \alpha} \right) \right) \right]\nonumber \\
    &=&\mathcal{F}\left[aL\left(\vec{\alpha}\right)\mathcal{F}\left[\mathrm{sinc}\left(x-\Theta\right)\right]-aL\left(\vec{\alpha}\right)\mathrm{sinc}\left(\Theta\right)\right]\nonumber \\&=&a\left(W_L\mathrm{sinc}\left(W_L \left(x+\Theta\right)\right)-\mathrm{sinc}\left(\Theta\right)W_L\mathrm{sinc}\left(W_L x\right)\right)\mathrm{sinc}\left(y\right),\label{c15}
\end{eqnarray}
where the symbol $W_L$ denotes the full width of the Lyot stop normalized by the pupil aperture full width.
Because the right-hand side of Equation \ref{c15} takes real-number values, it can have only two patterns of the phases: 0 rad (for positive values) and $\pi$ rad (for negative values). 
Hence, we can justify the assumption that phases of output focal amplitudes take constant values within their main lobes.
\subsection{Assessment of the Effect of Side Lobes}
In Equation (\ref{kkkk}), we ignored the negative values of the side lobes of the amplitude $S\left(\vec{x}\right)$ and assumed that $S\left(\vec{x}\right)=\left|S\left(\vec{x}\right)\right|$.
In this subsection, we evaluate the impact of this simplification on the resultant fiber-coupling efficiency.
\subsubsection{Theoretical Fiber-Coupling Efficiency for the Source with Separation Angles of 1$\lambda/D$}
First, we see the theoretical values of fiber-coupling efficiency without simplification below.
Using Equation (\ref{c15}), we can obtain the following expression for the normalized amplitude spread function $S\left(\vec{x}\right)$ in the case with the 1-$\lambda/D$ separated source:
\begin{eqnarray}
    S\left(\vec{x}\right)&=&\frac{a W_L\mathrm{sinc}\left(W_L \left(x+1\right)\right)\mathrm{sinc}\left(y\right)}{\sqrt{\int_{\mathbb{R}^2}d\vec{x} \left| a W_L\mathrm{sinc}\left(W_L \left(x+1\right)\right)\mathrm{sinc}\left(y\right) \right|^2}}\nonumber\\
    &=&\sqrt{W_L}\mathrm{sinc}\left(W_L \left(x+1\right)\right)\mathrm{sinc}\left(y\right). \label{c16}
\end{eqnarray}
Since locating the fiber center at the position $\left(x_c,y_c\right)=\left(-1,0\right)$ yields the highest coupling efficiency $\eta$, we assume the fiber-mode function $ g\left(\vec{x}\right)$as follows (see also Equation (\ref{poiu})):
\begin{equation}
 g\left(\vec{x}\right)=\frac{1}{\sqrt{2\pi} \sigma}e^{-\frac{\left(x+1\right)^2+y^2}{4\sigma^2}}. \label{c17}  
\end{equation}
Using Equations (\ref{c16}) and (\ref{c17}), we can calculate the coupling efficiency $\eta$ as the following:
\begin{eqnarray}
\eta&=&\left|\int_{\mathbb{R}^2}\!\! d\vec{x} \ g\left(\vec{x}\right)^{\ast} S\left(\vec{x}\right)\right|^2\nonumber \\
    &=&\frac{W_L}{2\pi \sigma^2}\left(\int_{-\infty}^{\infty}dx e^{-\frac{\left(x+1\right)^2}{4\sigma^2}} \mathrm{sinc}\left(W_L \left(x+1\right)\right)\right)^2\left(\int_{-\infty}^{\infty}dye^{-\frac{y^2}{4\sigma^2}} \mathrm{sinc}\left(y \right) \right)^2 \nonumber \\
   &=& \frac{W_L}{2\pi \sigma^2}\left(\int_{-\infty}^{\infty}dx e^{-\frac{x^2}{4\sigma^2}} \mathrm{sinc}\left(W_L x\right)\right)^2\left(\int_{-\infty}^{\infty}dye^{-\frac{y^2}{4\sigma^2}} \mathrm{sinc}\left(y \right) \right)^2.\label{c18}
\end{eqnarray}
To simplify Equation (\ref{c18}), we use the fact that the Fourier transform keeps the inner products as follows:
\begin{eqnarray}
\eta&=& \frac{W_L}{2\pi \sigma^2}\left(\int_{-\infty}^{\infty}d\alpha  \mathcal{F}\left[e^{-\frac{x^2}{4\sigma^2}}\right]  \mathcal{F}\left[\mathrm{sinc}\left(W_L x\right)\right]\right)^2\left(\int_{-\infty}^{\infty}d\beta  \mathcal{F}\left[e^{-\frac{y^2}{4\sigma^2}}\right]  \mathcal{F}\left[ \mathrm{sinc}\left(y \right)\right] \right)^2.\label{c18}
\end{eqnarray}
The following mathematical formula works here:
\begin{equation}
    \mathcal{F}\left[e^{-\frac{x^2}{4\sigma^2}}\right]=2\sigma \sqrt{\pi}e^{-4\sigma^2 \pi^2 \alpha^2}\label{c19}
\end{equation}
and
\begin{equation}
    \mathcal{F}\left[\mathrm{sinc}\left(W_L x\right)\right]=\frac{1}{W_L}\mathrm{rect}\left[\frac{\alpha}{W_L}\right].\label{c20}
\end{equation}
Substituting Equations (\ref{c19}) and (\ref{c20}) into Equation (\ref{c18}) leads to the following expression:
\begin{eqnarray}
\eta&=& \frac{W_L}{2\pi \sigma^2}\left(\int_{-\infty}^{\infty}d\alpha  2\sigma \sqrt{\pi}e^{-4\sigma^2 \pi^2 \alpha^2} \frac{1}{W_L}\mathrm{rect}\left[\frac{\alpha}{W_L}\right]\right)^2\left(\int_{-\infty}^{\infty}d\beta  2\sigma \sqrt{\pi}e^{-4\sigma^2 \pi^2 \beta^2}  \mathrm{rect}\left[\beta\right] \right)^2\nonumber\\
&=&\frac{W_L}{2\pi \sigma^2}\left(\int_{-\frac{W_L}{2}}^{\frac{W_L}{2}}d\alpha  \frac{2\sigma \sqrt{\pi}}{W_L}e^{-4\sigma^2 \pi^2 \alpha^2} \right)^2\left(\int_{-\frac{1}{2}}^{\frac{1}{2}}d\beta  2\sigma \sqrt{\pi}e^{-4\sigma^2 \pi^2 \beta^2} \right)^2 \nonumber \\
&=&\frac{W_L}{2\pi \sigma^2}\frac{1}{\pi^2 W_L^2}\left(\int_{-\sigma \pi W_L}^{\sigma \pi W_L}d\left(2\sigma \pi \alpha\right)  e^{-\left(2\sigma \pi \alpha\right)^2} \right)^2\left(\int_{-\sigma \pi }^{\sigma \pi }d\left(2\sigma \pi \beta\right)  e^{-\left(2\sigma \pi \beta\right)^2} \right)^2\nonumber\\
&=&\frac{1}{2\pi W_L }\frac{\mathrm{Erf}\left(W_L \pi \sigma \right)^2 \mathrm{Erf}\left( \pi \sigma \right)^2}{\sigma^2},\label{c22}
\end{eqnarray}
where $\mathrm{Erf}\left(z\right)$ denotes the error function defined as follows:
\begin{equation}
    \mathrm{Erf}\left(z\right)=\frac{2}{\sqrt{\pi}}\int_{0}^{z} dt\ e^{-t^2}.
\end{equation}
When $W_L=1.0$, the fiber-coupling efficiency $\eta(\sigma)$ as a function of the parameter $\sigma$ takes a maximum value of  about 79\% at $\sigma\sim 0.32$ according to the numerical evaluation.
When $W_L=0.75$, it takes a maximum value of about 77\% at $\sigma\sim 0.36$.
\subsubsection{Fiber-Coupling Efficiency by Ignoring the Phase of the Side Lobe}
Here, we define a numerical sequence $a_n$ $\left(n=1,2,3,...\right)$ as follows: 
\begin{equation}
    a_n \left(\sigma, W_L\right)=\int_{\frac{n-1}{W_L}}^{\frac{n}{W_L}} dx \ e^{-\frac{x^2}{4\sigma^2}} \mathrm{sinc}\left(W_L x\right). \label{c24}
\end{equation}
Using Equation (\ref{c24}), we can rewrite Equation (\ref{c18}):
\begin{equation}
    \eta=\frac{W_L}{2\pi \sigma^2}\left(2\sum^{\infty}_{n=0}a_n(\sigma,W_L)\right)^2\left(2\sum^{\infty}_{m=0}a_m(\sigma,1)\right)^2.
\end{equation}
Table \ref{tab1} shows the numerical values of $a_n$ ($n=1,2,3$) in the cases where $\left(\sigma, W_L\right)=\left(0.36, 0.75\right)$ or $\left(\sigma, W_L\right)=\left(0.36, 1.0\right)$; when Lyot-stop width $W_L$ takes 0.75, the parameter $\sigma$ of 0.36 leads to the maximum value of the coupling efficiency in Equation (\ref{c22}). 
\begin{table}[htb]
    \centering
    \begin{tabular}{c| c |c}\hline\hline
        $n$ & $a_n \left(\sigma=0.36, W_{L}=0.75\right)$ & $a_n \left(\sigma=0.36, W_{L}=1.0\right)$ \\ \hline
         1  & $5.1\times 10^{-1}$ & $5.1\times 10^{-1}$ \\
         2  &$-5.0\times 10^{-4}$ &$-3.7\times 10^{-3}$ \\
         3  & $3.2\times 10^{-9}$ & $2.6\times 10^{-6}$ \\
        $n+1$  & $\left|a_{n+1}\right|<\left|a_n\right| $ & $\left|a_{n+1}\right|<\left|a_n\right| $ \\
    \end{tabular}
    \caption{The numerical values of $a_n$ ($n=1,2,3$) in the cases where $\left(\sigma, W_L\right)=\left(0.36, 0.75\right)$ or $\left(\sigma, W_L\right)=\left(0.36, 1.0\right)$.}
    \label{tab1}
\end{table}
From the values in Tabel \ref{tab1}, we can conclude that ignoring the terms of $a_n$ where $2\leq n$ provides the coupling efficiency evaluation with few errors (error $\sim 1\%$).
Hence, the following equation approximately holds:
\begin{eqnarray}
    \eta&\sim& \frac{W_L}{2\pi \sigma^2}\left(2a_1(0.36,0.75)\right)^2\left(2a_1(0.36,1)\right)^2 \nonumber\\
       &\sim&\frac{W_L}{2\pi \sigma^2}\left(2\sum^{\infty}_{n=0}(-1)^{n-1}a_n(0.36,0.75)\right)^2\left(2\sum^{\infty}_{m=0}(-1)^{m-1}a_m(0.36,1)\right)^2\nonumber\\
        &=&\frac{W_L}{2\pi \sigma^2}\left(2\sum^{\infty}_{n=0}\left|a_n(0.36,0.75)\right|\right)^2\left(2\sum^{\infty}_{m=0}\left|a_m(0.36,1)\right|\right)^2\nonumber\\
    &=&\eta',
\end{eqnarray}
where the symbol $\eta'$ indicates the coupling efficiency evaluated under the assumption of Equation (\ref{kkkk}).
\end{document}